\documentclass[%
 reprint,
superscriptaddress,
 amsmath,amssymb,
 aps,
]{revtex4-1}

\usepackage{graphicx}
\usepackage{dcolumn}
\usepackage{bm}
\usepackage{subfigure}

\begin{document}

\preprint{APS/123-QED}

\title{Continuous-variable quantum key distribution \\ with non-Gaussian quantum catalysis}

\author{Ying Guo}%
\affiliation{School of Information Science and Engineering, Central South University, Changsha 410083, China}

\author{Wei Ye}%
\affiliation{School of Information Science and Engineering, Central South University, Changsha 410083, China}

\author{Hai Zhong}%
\affiliation{School of Information Science and Engineering, Central South University, Changsha 410083, China}

\author{Qin Liao}
\email{Corresponding author: llqqlq@csu.edu.cn}\affiliation{School of Information Science and Engineering, Central South University, Changsha 410083, China}
\affiliation{School of Electrical and Electronic Engineering, Nanyang Technological University, Singapore 639798, Singapore}

\date{\today}

\begin{abstract}
The non-Gaussian operation can be used not only to enhance and distill the
entanglement between Gaussian entangled states, but also to improve quantum
communications. In this paper, we propose an non-Gaussian
continuous-variable quantum key distribution (CVQKD) by using quantum
catalysis (QC), which is an intriguing non-Gaussian operation in essence
that can be implemented with current technologies. We perform quantum
catalysis on both ends of the Einstein-Podolsky-Rosen (EPR) pair prepared by
a sender, Alice, and find that for the single-photon QC-CVQKD, the bilateral
symmetrical quantum catalysis (BSQC) performs better than the single-side
quantum catalysis (SSQC). Attributing to characteristics of integral within
an ordered product (IWOP) of operators, we find that the quantum catalysis
operation can improve the entanglement property of Gaussian entangled states
by enhancing the success probability of non-Gaussian operation, leading to
the improvement of the QC-CVQKD system. As a comparison, the QC-CVQKD system
involving zero-photon and single-photon quantum catalysis outperforms the
previous non-Gaussian CVQKD scheme via photon subtraction in terms of secret
key rate, maximal transmission distance and tolerable excess noise.

\end{abstract}
\maketitle

\section{Introduction}

Quantum key distribution (QKD) \cite{1,2,3,4,5}, as one of the mature
practical applications in quantum information processing, allows two distant
legitimate parties (normally say a sender, Alice and a receiver, Bob) to
establish a set of secure keys even in the presence of the untrusted
environment controlled by an eavesdropper (Eve), and its unconditional
security can be guaranteed by the laws of quantum physics, e.g., the
uncertainty principle \cite{6} and the non-cloning theorem \cite{7}. In
general, QKD mainly includes two families, namely discrete-variable quantum
key distribution (DVQKD) and continuous-variable quantum key distribution
(CVQKD) \cite{2,8,9,10,11,12,13,14,15}. Due to the limitation of
single-photon detectors employed in DVQKD systems, in CVQKD, the sender
Alice encodes information on the quadratures of the optical field with
Gaussian modulation and the receiver Bob decodes the secret information with
high-speed and high-efficiency homodyne or heterodyne detection so that
CVQKD has become the center of attention in recent years \cite{16,17}. In
addition, since the security proofs of the Gaussian-modulated CVQKD
protocols against collective attacks \cite{16,18} and coherent attacks \cite%
{19,20} have been proven experimentally \cite{21,22,23}, the
Gaussian-modulated CVQKD protocols take on the potential application
prospects of long-distance communication. Among them, the Grosshans-Grangier
2002 (GG02) protocol \cite{17} performs outstandingly over short distance,
but seems unfortunately to be facing the problem of long-secure distance
compared with its DVQKD counterpart.

Till now, many remarkable theoretical and experimental efforts have been
devoted to extending the maximal transmission distance with high rate in
CVQKD systems \cite{23,24,25,26,27,28}. By the use of multidimensional
reconciliation protocols in the regime of low signal-to-noise ratio (SNR)
\cite{23}, it was demonstrated experimentally that CVQKD over $80$ km
transmission distance can be realized. The reason is that the
multidimensional reconciliation is, in a sense, to design a suitable
reconciliation code with high efficiency even at low SNR, which can increase
the secure distance \cite{28}. Alternatively, the discrete modulation
protocols such as the four-state protocol \cite{13,27,29,30} and eight-state
protocol \cite{31} were shown to improve the secure distance as there does
exist suitable error-correlation codes with high efficiency for discrete
possible values at low SNR. Especially for eight state protocol, not only
can the secret key rate be improved, but the transmission distance more than
$100$ km can be achieved \cite{31,32}. From a practical point of view, the
maximal transmission distance and the unconditional security of the secret
key are usually disturbed by the environmental noise and dissipation. To
solve these problems,\ the methods of source monitoring \cite{33} and linear
optics cloning machine \cite{34} have been proposed subsequently.

Thanks to the development of experimental techniques, on the other hand,
some quantum operations have been used to improve the performance of CVQKD
in terms of secret key rate and tolerable excess noise. For example, a
heralded noiseless amplifier \cite{26,27,35} was proposed to improve the
maximal transmission distance roughly by the equivalent of 20log$_{\text{10}%
} $g dB losses resulting from the compensation of the losses \cite{27}.
Recently, due to the fact that the non-Gaussian operation can be used for
improving the entanglement \cite{36,37,38} and quantum teleportation in CV
system \cite{39,40}, the photon-subtraction operation, which is one of the
non-Gaussian operations, has been proposed to improve the secret key rate,
the maximal tolerable excess noise and the transmission distance of CVQKD
protocol \cite{11,14,15,29}. In particular, the single-photon subtraction
operation in the enhanced CVQKD protocol\ outperforms other numbers of
photon subtraction. Unfortunately, the success probability for implementing
this single-photon subtraction operation at the variance of two-mode
squeezed vacuum (TMSV) state $V=20\ $is limited to below $0.25$, which may
lead to loss more information between Alice and Bob in the process of
extracting the secret key. In order to overcome the limitation, in this
paper, we propose an improved modulation scheme for CVQKD by using another
non-Gaussian operation, the quantum catalysis (QC) \cite{41}, which can be
implemented with current experimental technologies. Attractively, the
quantum catalysis operation is a feasible way to enhance the nonclassicality
\cite{42} and the entanglement property of Gaussian entangled states \cite%
{43}, thereby has become one of the research hotspots in quantum physics.
Different from the previous studied photon-subtraction operations, although
no photon is subtracted and added in quantum catalysis process, quantum
catalysis can be applied to facilitate the conversion of the target
ensemble, which could prevent the loss of information effectively. Numerical
simulation shows that the entanglement and the success probability for
implementing quantum catalysis can be improved efficiently. Specifically,
the success probability for implementing zero-photon quantum catalysis can
be dramatically enhanced when compared with the previous CVQKD with
single-photon subtraction. In addition, we illustrate the performance of
QC-CVQKD with different photon-catalyzed numbers, and find that zero-photon
and single-photon catalysis presents the well performance when optimized
over the transmittance $T$ of Alice$^{\text{'}}$s beam splitter (BS).

This paper is organized as follows. In Sec.II, we suggest a quantum
catalysis operator, and detail the process of QC-CVQKD. In Sec.III, the
success probability and the entanglement property for implementing quantum
catalysis are analyzed, and security analysis for QC-CVQKD system is
subsequently discussed. Finally, conclusions are drawn in Sec.IV.

\section{Quantum catalysis-based CVQKD}

To make the derivation self-contained, we suggest quantum catalysis operator
by using the technique of integral within an ordered product (IWOP) of
operators \cite{44}, and then detail the QC-CVQKD.

\subsection{Quantum catalysis operation}

\begin{figure}[tbp]
\label{Fig.1} \centering \includegraphics[width=0.85\columnwidth]{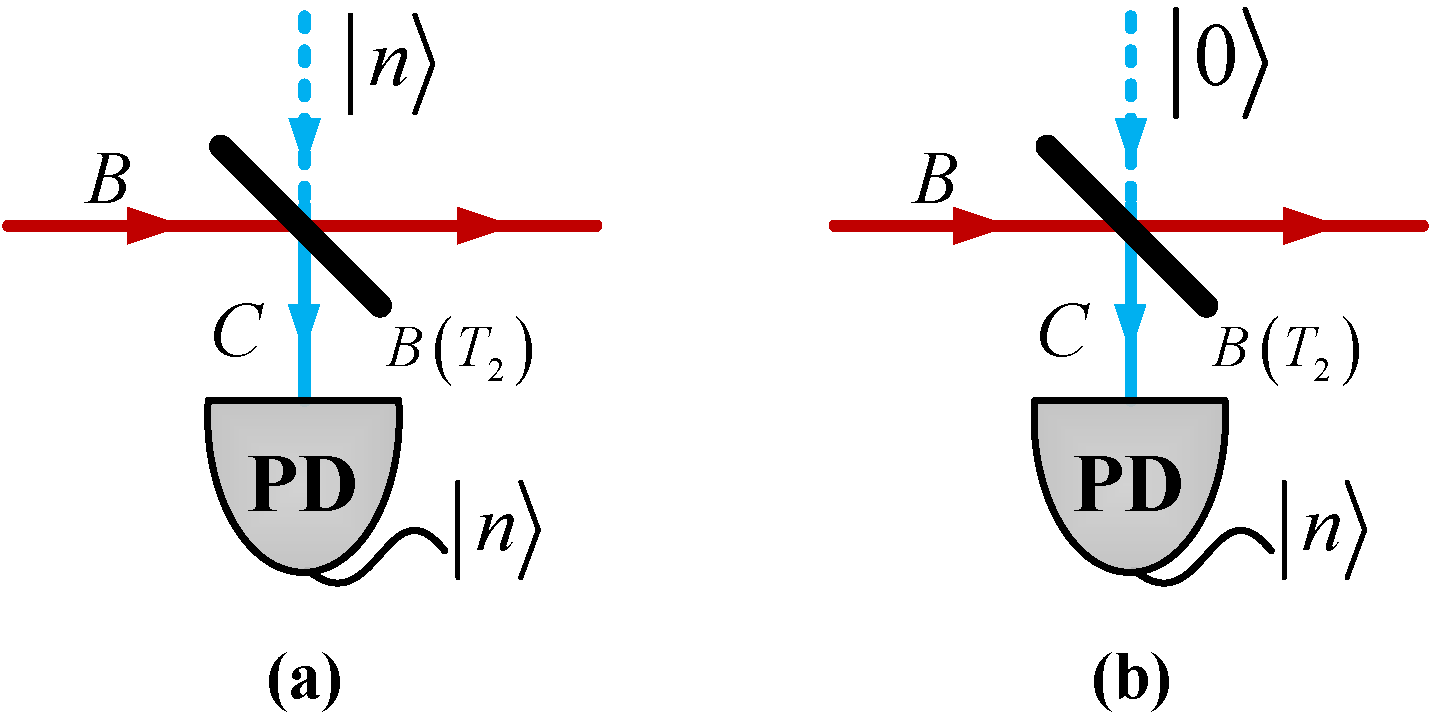}
\caption{ (Color online) Schematic diagram of the non-Gaussian operations.
(a) The quantum catalysis (QC). An n-photon Fock state $\left\vert
n\right\rangle $ in auxiliary mode C is split on the asymmetrical beam
splitter (BS) with transmittance $T_{2}$. Subsequently, a photon detector
(PD) at the auxiliary mode is performed by the conditional detection of $%
\left\vert n\right\rangle $, which is the so-called $n$-photon quantum
catalysis represented by an equivalent operator $\hat{O}_{n}$. (b) The $n$%
-photon-subtraction operation. A vacuum state $\left\vert 0\right\rangle $
in auxiliary mode C is injected into the asymmetrical beam splitter (BS)
with transmittance $T_{2}$. Likewise, a photon detector (PD) at the
auxiliary mode is performed by the conditional detection of $\left\vert
n\right\rangle .$}
\end{figure}

As shown in Fig.1(a), an $n$-photon Fock state $\left \vert n\right \rangle $
in auxiliary mode C is injected at one of the input ports of BS with
transmittance $T_{2}$, and simultaneously detected at the corresponding
output port of BS, which is the so-called $n$-photon catalysis. In fact,
this quantum catalysis process is often regarded as an equivalent operator $%
\hat{O}_{n}$ given by%
\begin{equation}
\hat{O}_{n}\equiv_{C}\left \langle n\right \vert B\left( T_{2}\right) \left
\vert n\right \rangle _{C},  \label{1}
\end{equation}
where $B\left( T_{2}\right) $ is the BS operator with transmittance $T_{2}$.
To obtain the specific expression of the equivalent operator $\hat{O}_{n}$,
we employ the normally order form of $B\left( T_{2}\right) $ by the IWOP
technique and the coherent state representation of Fock state $\left \vert
n\right \rangle $, which are respectively expressed as $B\left( T\right) =:%
\exp[(\sqrt{T_{2}}-1)(b^{\dag}b+c^{\dag}c)+\left( c^{\dag
}b-cb^{\dagger}\right) \sqrt{1-T_{2}}]:$ and $\left \vert n\right \rangle =1/%
\sqrt{n!}\frac{\partial^{n}}{\partial \beta^{n}}\left \Vert \beta
\right
\rangle |_{\beta=0}$ where the notations :$\cdot$: and $\left \Vert
\beta \right \rangle =\exp(\beta c^{\dagger})\left \vert 0\right \rangle $
represent a normally ordering of operator and an un-normalized coherent
state, respectively. As a result, Eq.(\ref{1}) can be described as
\begin{equation}
\hat{O}_{n}=:L_{n}\left( \frac{1-T_{2}}{T_{2}}b^{\dagger}b\right) :\left(
\sqrt{T_{2}}\right) ^{b^{\dagger}b+n},  \label{2}
\end{equation}
where $L_{n}\left( \cdot \right) $ denotes the Laguerre polynomials (see
Refs.\cite{42,43} for the detailed calculation). By using the generating
function of Laguerre polynomials, i.e.,%
\begin{equation}
L_{n}\left( x\right) =\frac{\partial^{n}}{n!\partial \gamma^{n}}\left \{
\frac{e^{\frac{-x\gamma}{1-\gamma}}}{1-\gamma}\right \} _{\gamma=0},
\label{3}
\end{equation}
and the operator relation $e^{\lambda b^{\dagger}b}=:\exp \left \{ \left(
e^{\lambda}-1\right) b^{\dagger}b\right \} :$, Eq.(\ref{2})\ can be further
rewritten as\
\begin{equation}
\hat{O}_{n}=G_{T_{2}}\left( b^{\dagger}b\right) \left( \sqrt{T_{2}}\right)
^{b^{\dagger}b+n},  \label{4}
\end{equation}
where \ \
\begin{equation}
G_{T_{2}}\left( b^{\dagger}b\right) =\frac{\partial^{n}}{n!\partial
\gamma^{n}}\left \{ \frac{1}{1-\gamma}\left( \frac{1-\gamma/T_{2}}{1-\gamma }%
\right) ^{b^{\dagger}b}\right \} _{\gamma=0}.  \label{5}
\end{equation}

From Eq.(\ref{4}), we find that the quantum catalysis operation belongs to a
kind of non-Gaussian operation. Moreover, as shown in Fig.1(a), for an
arbitrary input state $\left\vert \varphi \right\rangle _{in}$ in mode B,
the output state $\left\vert \psi \right\rangle _{out}$ can be expressed as $%
\left\vert \psi \right\rangle _{out}=$\ $\hat{O}_{n}$/$\sqrt{p}\left\vert
\varphi \right\rangle _{in}$ with the success probability $p$ for
implementing the $n$-photon catalysis $\hat{O}_{n}$, which is beneficial for
calculating the analytical expressions of the Alice output state and the the
covariance matrix between Alice and Bob in the following. In addition,
different from the $n$-photon-subtraction operation shown in Fig.1(b), no
photon is subtracted or added in $n$-photon catalysis operation. Such
operation facilitates the transformation between input and output states,
thereby effectively preventing useful information from being lost. However,
no matter how many photons are catalyzed or subtracted, there is no
quantum-catalysis or photon-subtraction effect when $T_{2}=1$.

\begin{figure*}[ptb]
\label{Fig.2} \centering \includegraphics[width=0.8\linewidth]{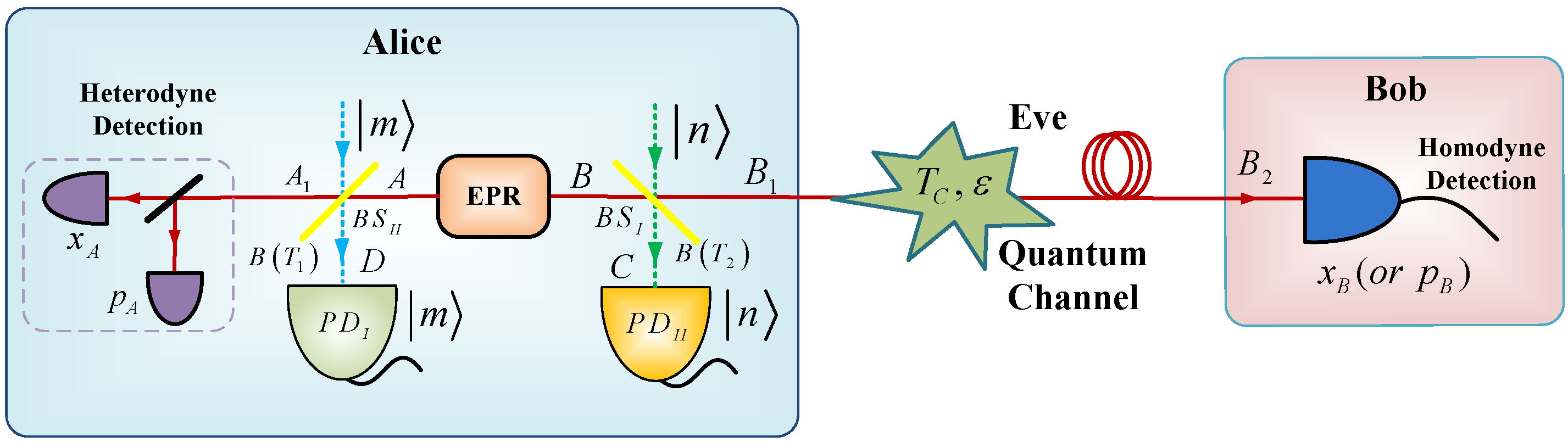}
\caption{(Color online) Schematic diagram of QC-CVQKD. EPR: two-mode
squeezed vacuum state. $\left \vert m\right \rangle $\ and $\left \vert
n\right \rangle $: $m$-photon and $n$-photon Fock state. PD$_{\text{I}}$ and
PD$_{\text{II}}$: photon detector by conditional measurement of $\left \vert
m\right \rangle $\ and $\left \vert n\right \rangle $. $B\left( T_{1}\right)
$ and $B\left( T_{2}\right) $: BS$_{\text{I}}$ operator with transmittance $%
T_{1}$ and BS$_{\text{II}}$ operator with transmittance $T_{2}$. $T_{c}$ and
$\protect\varepsilon$: quantum channel parameters.}
\end{figure*}
\ \ \ \ \ \ \ \ \ \ \ \ \ \ \ \ \ \ \ \ \ \ \ \ \ \ \ \ \ \ \ \ \ \ \ \ \ \
\ \ \ \ \ \ \ \ \ \ \ \ \ \ \ \ \ \ \ \ \ \ \ \ \ \ \ \ \ \ \ \ \ \ \ \ \ \
\ \ \ \ \ \ \ \ \ \ \ \ \ \ \ \ \ \ \ \ \

\subsection{The CVQKD protocol with quantum catalysis}

In what follows, we elaborate the schematic diagram of the QC-CVQKD
protocol, as shown in Fig.2. The sender, Alice generates a TMSV state (which
is also called as an Einstein-Podolsky-Rosen (EPR) pair) involving two modes
A and B with a modulation variance $V$, which is usually expressed as the
two-mode squeezed operator $S_{2}\left( r\right) =\exp \left \{ r\left(
a^{\dagger}b^{\dag}-ab\right) \right \} $ on the two-mode vacuum state $%
\left \vert 0,0\right \rangle _{AB}$, i.e.,

\begin{align}
\left\vert TMSV\right\rangle _{AB}& =S_{2}\left( r\right) \left\vert
0,0\right\rangle _{AB}  \notag \\
& =\sqrt{1-\lambda ^{2}}\underset{l=0}{\overset{\infty }{\sum }}\lambda
^{l}\left\vert l,l\right\rangle _{AB},  \label{6}
\end{align}%
where $\lambda =\tanh r=\sqrt{\left( V-1\right) /\left( V+1\right) }$ for $%
V=2\alpha ^{2}+1$, and $\left\vert l,l\right\rangle _{AB}=$ $\left\vert
l\right\rangle _{A}\otimes \left\vert l\right\rangle _{B}$ denotes the
two-mode Fock state of both modes A and B. After that, she performs $m$%
-photon and $n$-photon catalysis operations in modes A and B, respectively,
giving birth to the state $\left\vert \psi \right\rangle _{A_{1}B_{1}}$.
Note that, in mode A, inserting another quantum catalysis operation $\hat{O}%
_{m}$ at the end of the transmission to Alice is designed to figure out what
effect quantum catalysis has on the information between Alice and Bob, when
comparing with the single-side quantum catalysis $\hat{O}_{n}$ case.
According to the afore-mentioned method of quantum catalysis operation,
likewise in Eq.(\ref{4}), we obtain the $m$-photon quantum catalysis
operation, i.e.,%
\begin{equation}
\hat{O}_{m}=G_{T_{1}}\left( a^{\dagger }a\right) \left( \sqrt{T_{1}}\right)
^{a^{\dagger }a+m},  \label{7}
\end{equation}%
with the notation $G_{T_{1}}\left( a^{\dagger }a\right) $ given by
\begin{equation}
G_{T_{1}}\left( a^{\dagger }a\right) =\frac{\partial ^{m}}{m!\partial \tau
^{m}}\left\{ \frac{1}{1-\tau }\left( \frac{1-\tau /T_{1}}{1-\tau }\right)
^{a^{\dagger }a}\right\} _{\tau =0}.  \label{8}
\end{equation}%
Then, the yielded state $\left\vert \psi \right\rangle _{A_{1}B_{1}}$ turns
out to be%
\begin{align}
\left\vert \psi \right\rangle _{A_{1}B_{1}}& =\frac{\hat{O}_{m}\hat{O}_{n}}{%
\sqrt{P_{d}}}\left\vert TMSV\right\rangle  \notag \\
& \!\!\!\!\!\!\!\!\!\!\!\!\!\!\!\!=\underset{l=0}{\overset{\infty }{\sum }}%
\frac{W_{0}}{\sqrt{P_{d}}}\frac{\partial ^{m}}{\partial \tau ^{m}}\frac{%
\partial ^{n}}{\partial \gamma ^{n}}\frac{W^{l}}{(1-\tau )(1-\gamma )}%
\left\vert l,l\right\rangle _{AB},  \label{9}
\end{align}%
where $P_{d}$ denotes the success probability of implementing quantum
catalysis, which is an important indicator that affects the mutual
information in the process of distilling a common secret key between Alice
and Bob, and\ can be calculated as
\begin{equation}
P_{d}=W_{0}^{2}\Re ^{m,n}\left\{ \frac{\Pi }{1-W_{1}W}\right\} ,  \label{10}
\end{equation}%
with $\Re ^{m,n},\Pi ,W_{0},W$ and $W_{1}$ defined in Eq.(A2). Detailed
calculations of the success probability $P_{d}$ can be shown in Appendix A.
From Eq.(\ref{9}), the state $\left\vert \psi \right\rangle _{A_{1}B_{1}}$
becomes a non-Gaussian entangled state.

At Alice$^{\text{'}}$s station, the quadratures of both $x_{A}$ and $p_{A}$
are measured via heterodyne detection on the incoming one half of the state $%
\left \vert \psi \right \rangle _{A_{1}B_{1}}$, and the other half of $%
\left
\vert \psi \right \rangle _{A_{1}B_{1}}$ is sent to Bob through an
insecure quantum channel that can be controlled by Eve with the transmission
efficiency $T_{c}$ and the excess noise $\varepsilon$. After receiving the
state, Bob randomly chooses to measure either $x_{B}$ or $p_{B}$ via
homodyne detection and informs Alice about the measured observable. Finally,
Alice and Bob can share a string of secret keys by data-postprocessing.

Before deriving the performance of the QC-CVQKD protocol, we demonstrate the
entanglement of both the Gaussian entangled state $\left\vert
TMSV\right\rangle _{AB}$ and the transformed state $\left\vert \psi
\right\rangle _{A_{1}B_{1}}$. As a computable measurement of entanglement
and an upper bound on the distillable entanglement, the logarithmic
negativity is usually used to quantify the degree of entanglement, which is
given by%
\begin{equation}
E_{N}=\log _{2}\left\Vert \rho ^{PT}\right\Vert ,  \label{11}
\end{equation}%
in which $\rho ^{PT}$ is the partial transpose of density operator $\rho $
about arbitrary subsystem, and the symbol $\left\Vert \cdot \right\Vert $ is
the trace norm. By using the Schmidt decomposition, if an arbitrary state $%
\left\vert \Psi \right\rangle $ can be decomposed as $\left\vert \Psi
\right\rangle =\sum_{n=0}^{\infty }w_{n}\left\vert n\right\rangle
_{A}\left\vert n^{\prime }\right\rangle _{B}$ with the positive real number $%
w_{n}$ and the orthonormal states $\left\vert n\right\rangle _{A}$ and $%
\left\vert n^{\prime }\right\rangle _{B}$, its logarithmic negativity can be
calculated as%
\begin{equation}
E_{N}=2\log _{2}\left\vert \overset{\infty }{\underset{n=0}{\sum }}%
w_{n}\right\vert .  \label{12}
\end{equation}%
According to Eqs.(\ref{6}) and (\ref{9}), the logarithmic negativity of both
the TMSV state \cite{14,15} and the resulted state $\left\vert \psi
\right\rangle _{A_{1}B_{1}}$ can be, respectively, calculated as
\begin{align}
& E_{N}\left( \left\vert TMSV\right\rangle _{AB}\right) =-\log _{2}\left(
1+\alpha ^{2}\right)  \notag \\
& \quad \quad \quad \quad \quad \quad \quad \quad \quad -2\log _{2}\left(
\sqrt{1+\alpha ^{2}}-\alpha \right) ,  \notag \\
& E_{N}\left( \left\vert \psi \right\rangle _{A_{1}B_{1}}\right) =2\log
_{2}\left\vert \overset{\infty }{\underset{n=0}{\sum }}\frac{W_{0}}{\sqrt{%
P_{d}}}\frac{\partial ^{m}}{\partial \tau ^{m}}\frac{\partial ^{n}}{\partial
\gamma ^{n}}\right.  \notag \\
& \quad \quad \quad \quad \quad \quad \quad \left. \times \frac{W^{l}}{%
\left( 1-\tau \right) \left( 1-\gamma \right) }\right\vert .  \label{13}
\end{align}

\section{Performance analysis}

In this section, we demonstrate the success probability regarding quantum
catalysis operation, and derive the performance of the QC-CVQKD system in
terms of secret key rate and tolerable excess noise. A performance
comparison between the QC-CVQKD and the photon-subtracted CVQKD is made to
highlight the merits of the QC-based system. Note that for a simple and
convenient discussion, we consider two special cases, i.e., the bilateral
symmetrical quantum catalysis (BSQC, in which $T_{1}=T_{2}=T$ and $m=n$) and
the single-side quantum catalysis (SSQC, in which $T_{1}=1,T_{2}=T$ and $n$).

\subsection{Success probability for quantum catalysis}

\begin{figure}[ptb]
\label{Fig.3} \centering \includegraphics[width=\columnwidth]{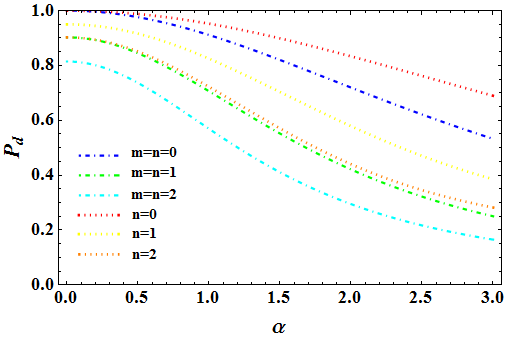}
\caption{(Color online) The success probability $P_{d}$ of quantum catalysis
as a function of $\protect\alpha$ for bilateral symmetrical quantum
catalysis (BSQC) ($T_{1}=T_{2}=T$ and $m=n\in\{0,1,2\}$) [dash-dotted line]
and single side quantum catalysis (SSQC) ($T_{1}=1$, $T_{2}=T$ and $%
n\in\{0,1,2\}$) [dotted line] for $T=0.95$. }
\end{figure}

The explicit form of the success probability for implementing quantum
catalysis operations has been given in Eq.(\ref{9}). In particular, for the
zero-photon BSQC ($T_{1}=T_{2}=T$ and $m=n=0$) and SSQC ($T_{1}=1,T_{2}=T$
and $n=0$), the success probabilities for implementing such zero-photon
quantum catalysis operations can be given by 1/(1-$\left( T^{2}-1\right)
\alpha ^{2}$)\ and 1/(1-$\left( T-1\right) \alpha ^{2}$), respectively.
Given a high transmittance $T=0.95$, the success probabilities $P_{d}$ can
be plotted as a function of $\alpha $ with several different
photon-catalysis numbers such as $m,n\in \{0,1,2\}$. Fig.3 shows that the
overall trend of success probability $P_{d}$ decreases as $\alpha $
increases. It indicates that for the increased modulation variance $%
V=2\alpha ^{2}+1$, the success probability $P_{d}$ of implementing quantum
catalysis decreases. Meanwhile, the success probabilities decrease with the
increased number of photon catalysis for both SSQC and SBQC. The
above-mentioned phenomenon explains that the implementation of multiphoton
catalysis ($m=n>1$ and $n>1$) may be relatively difficult to achieve.
Whereas, the success probability $P_{d}$ of SSQC provides better performance
than that of BSQC when one considers same photon-catalyzed numbers. For the
zero-photon SSQC ($n=0$) and BSQC ($m=n=0$), the success probabilities $%
P_{d} $ for the given large $\alpha $ $(\alpha =3)$ are approximate $0.68$
and $0.53$, respectively. It is worth noting that for the two-photon BSQC ($%
m=n=2$), the success probability $P_{d}$ for the given large $\alpha $ $%
(\alpha =3) $ is below 0.2, which may leak much information in the CVQKD
system.

Now we consider the effect of entanglement variation on the QC-CVQKD system,
which can be evaluated by the logarithmic negativity in Eq.(\ref{13}). For
arbitrary photon-catalyzed numbers $m$ and $n$, we can obtain the
logarithmic negativity of the state $\left\vert \psi \right\rangle
_{A_{1}B_{1}}$. Given a high transmittance $T=0.95$, we plot the logarithmic
negativity of both $E_{N}$($\left\vert \psi \right\rangle _{A_{1}B_{1}}$)
and $E_{N}$($\left\vert TMSV\right\rangle _{AB}$) as a function of $\alpha $
involving different photon-catalyzed numbers, as shown in Fig.4. For the
zero-photon and single-photon quantum catalysis, the entanglement property
can be improved for $\alpha =3$, which may have an important impact on the
correlation strength of mutual information between Alice and Bob. However,
for $\alpha =3$, the gap of the enhanced entanglement in BSQC decreases with
the increase of $m,n\in \{0,1,2\}$. The similar trend occurs to SSQC and
there is no improvement of the entanglement for $n=2$. Although the
entanglement for $m=n=2$ can be improved at large region of $\alpha $, there
does exist the limitation of its success probability. These results show
that the zero-photon and single-photon quantum catalysis (i.e, $m=n\in
\{0,1\}$ and $n\in \{0,1\}$) perform well in terms of the success
probability and the entanglement property, when comparing with the
two-photon cases (i.e, $m=n=2$ and $n=2$). On the other hand, for the
optimized $T$, we give the optimal logarithmic negativity $E_{N}$ as a
function of $\alpha $ for $m=n\in \{0,1\}$ and $n\in \{0,1\}$, as shown in
Fig.5. We find that the optimal entanglements of different zero-photon and
single-photon quantum catalysis cases overlap together, and then the gap of
the improved entanglement increases with the increasing $\alpha $.

To highlight the contribution of quantum catalysis operation, compared with
single-photon subtraction, we illustrate the success probability and the
entanglement property in Fig.6. For $T\rightarrow1,$ although the
improvement of the entanglement for single-photon subtraction [magenta
surface] performs better than that for quantum catalysis operation, the
success probability for the former is worse than that for the latter. As a
result, the quantum catalysis operation is superior to the single-photon
subtraction in terms of the success probability. These results indicate that
the quantum catalysis, as a novel non-Gaussian operation, can be used to
improve the entanglement property of Gaussian entangled states, and has an
advantage of the success probability over the photon subtraction operation.
Consequently, in what follows, we focus on quantum catalysis to enhance the
performance of the CVQKD system.

\begin{figure}[ptb]
\label{Fig.4} \centering \includegraphics[width=\columnwidth]{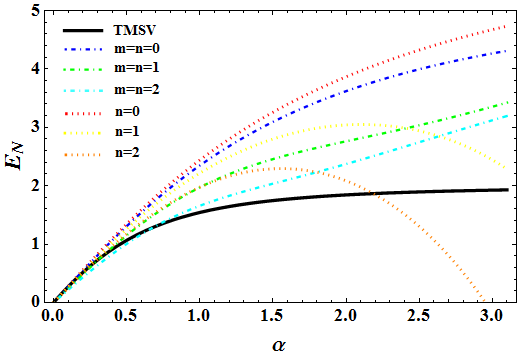}
\caption{(Color online) The logarithmic negativity of $E_{N}$($\left \vert
\protect\psi \right \rangle _{A_{1}B_{1}}$) as a function of $\protect\alpha$
for bilateral symmetrical quantum catalysis (BSQC) ($T_{1}=T_{2}=T$ and $%
m=n\in\{0,1,2\}$) [dash-dotted line] and single side quantum catalysis
(SSQC) ($T_{1}=1$, $T_{2}=T$ and $n\in\{0,1,2\}$) [dotted line] for $T=0.95$%
. }
\end{figure}
\begin{figure}[ptb]
\label{Fig.5} \centering \includegraphics[width=\columnwidth]{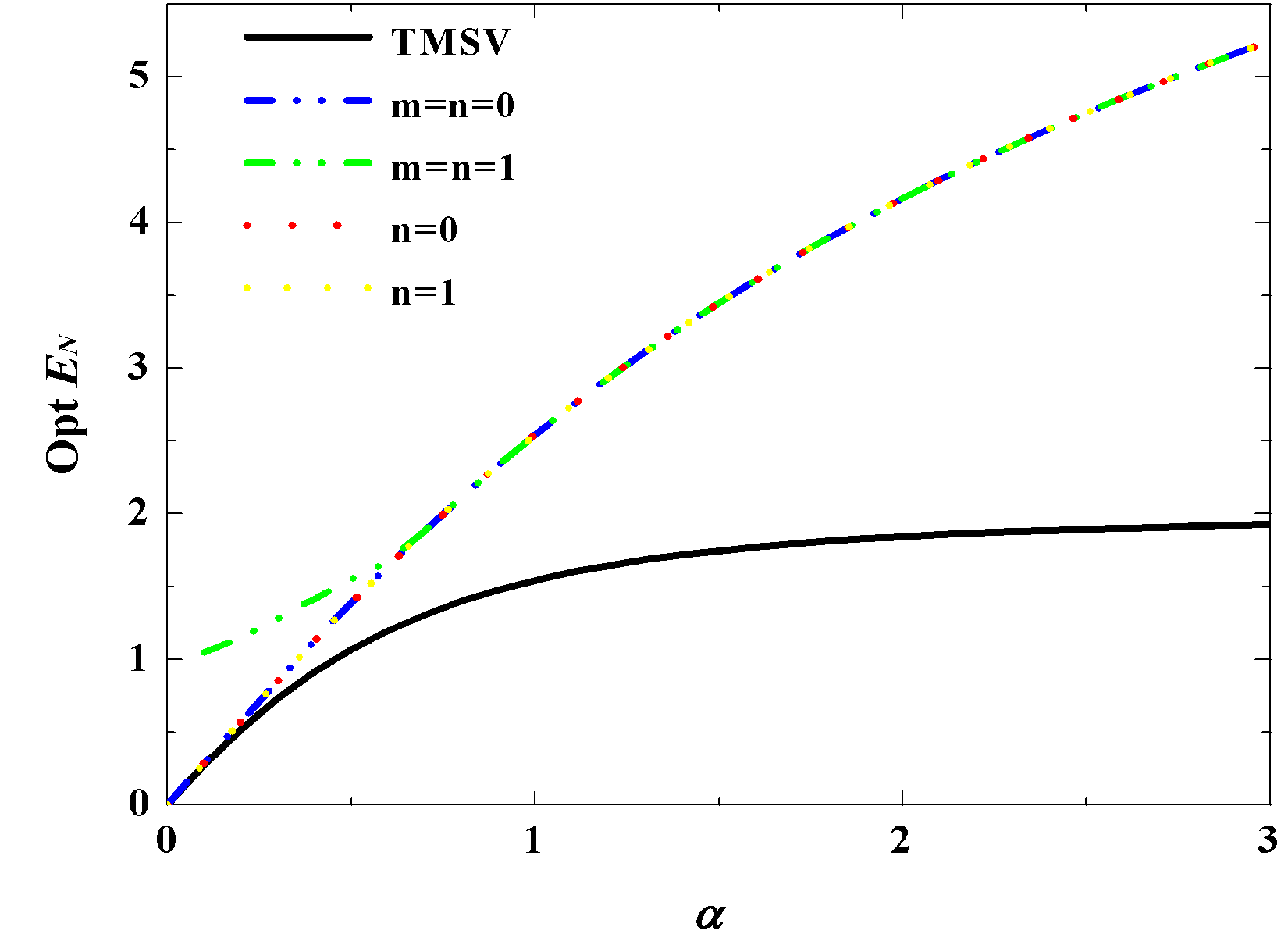}
\caption{(Color online) The optimal logarithmic negativity of $E_{N}$($%
\left
\vert \protect\psi \right \rangle _{A_{1}B_{1}}$) as a function of $%
\protect\alpha$ for bilateral symmetrical quantum catalysis (BSQC) ($%
T_{1}=T_{2}=T$ and $m=n\in\{0,1\}$) [dash-dotted line] and single side
quantum catalysis (SSQC)($T_{1}=1$, $T_{2}=T$ and $n\in\{0,1\}$) [dotted
line] for the optimal choice of $T$. \ \ }
\end{figure}
\begin{figure}[ptb]
\label{Fig.6} \centering \includegraphics[width=\columnwidth]{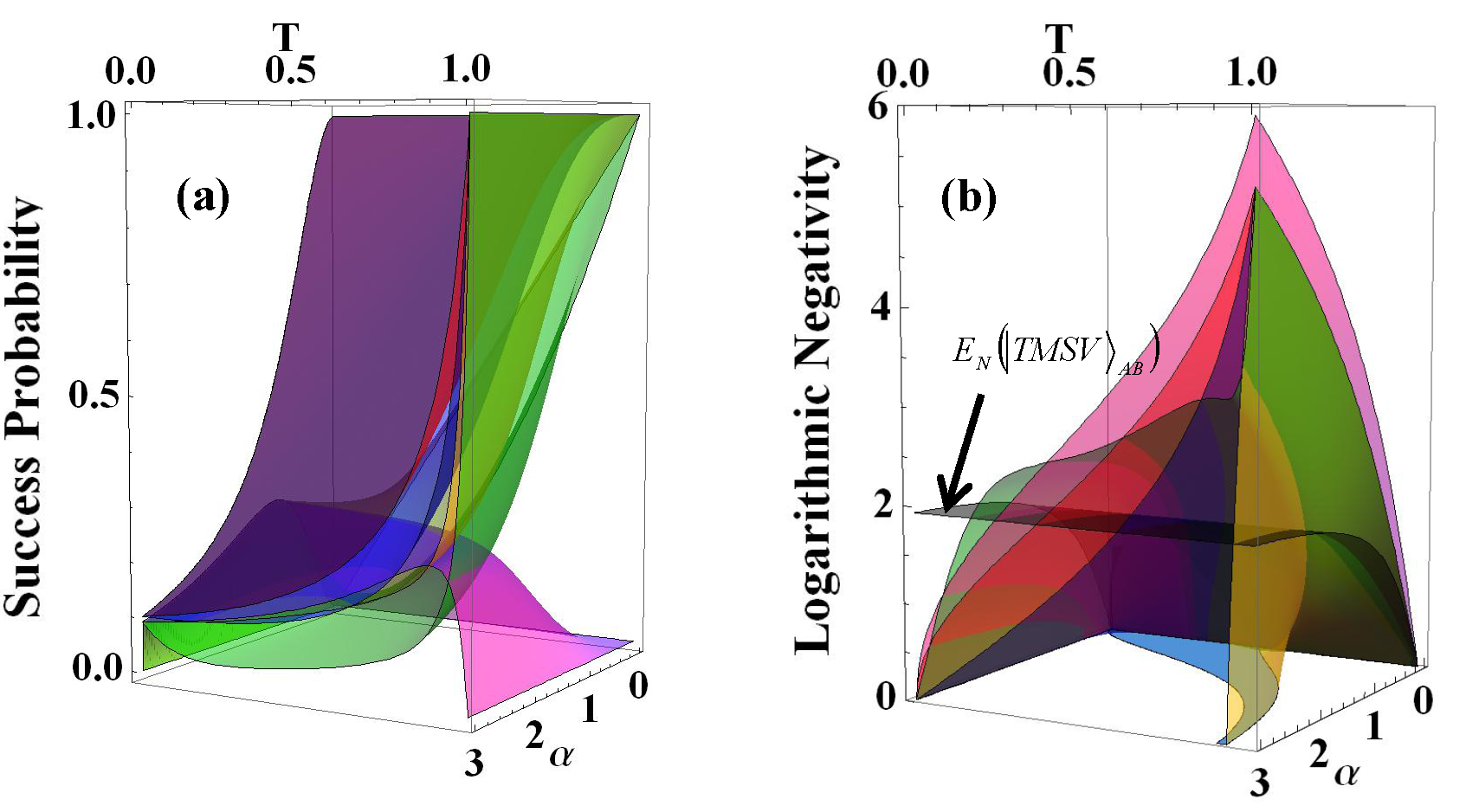}
\newline
\ \ \ \ \
\caption{(Color online) (a) The success probability of implementing between
quantum catalysis (QC) and single-photon subtraction in $\left( T,\protect%
\alpha \right) $ space with different photon-catalyzed numbers. (b) The
logarithmic negativity for the resulted state $\left \vert \protect\psi %
\right \rangle _{A_{1}B_{1}}$ using quantum catalysis (QC) and the
single-photon subtraction state $\left \vert \Psi \right \rangle _{AB_{1}}$
as well as the TMSV state $\left \vert TMSV\right \rangle _{AB}$ in $\left(
T,\protect\alpha \right) $ space with different photon-catalyzed numbers. In
(a) and (b), magenta surface stands for the single-photon subtraction case.
Other surfaces denote $m=n=0$ [blue surface], $m=n=1$ [green surface], $n=0$
[red surface] and $n=1$ [yellow surface].\ }
\end{figure}

\begin{figure}[ptb]
\label{Fig.7} \centering \includegraphics[width=\columnwidth]{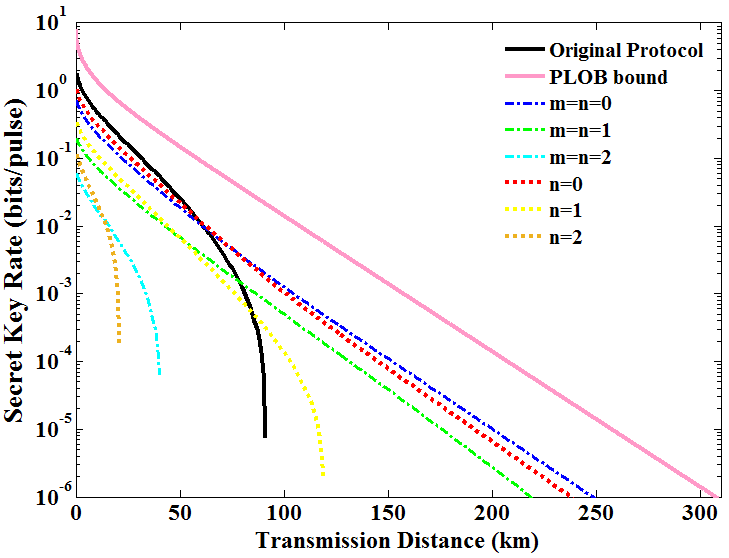}
\caption{(Color online) The asymptotic secret key rates of the QC-CVQKD
system [dash-dotted lines and dotted lines] as a function of the
transmission distance for bilateral symmetrical quantum catalysis (BSQC)
cases ($T_{1}=T_{2}=T$ and $m=n\in\{0,1,2\}$) [dash-dotted line] and single
side quantum catalysis (SSQC) cases ($T_{1}=1$, $T_{2}=T$ and $n\in\{0,1,2\}$%
) [dotted line] with $T=0.95$. The variance of EPR state is $V=20$, the
excess noise is $\protect\varepsilon=0.01$, and the reconciliation
efficiency is $\protect\beta=0.95$.}
\end{figure}
\begin{figure}[ptb]
\label{Fig.8} \centering
\subfigure[]{
\includegraphics[width=\columnwidth]{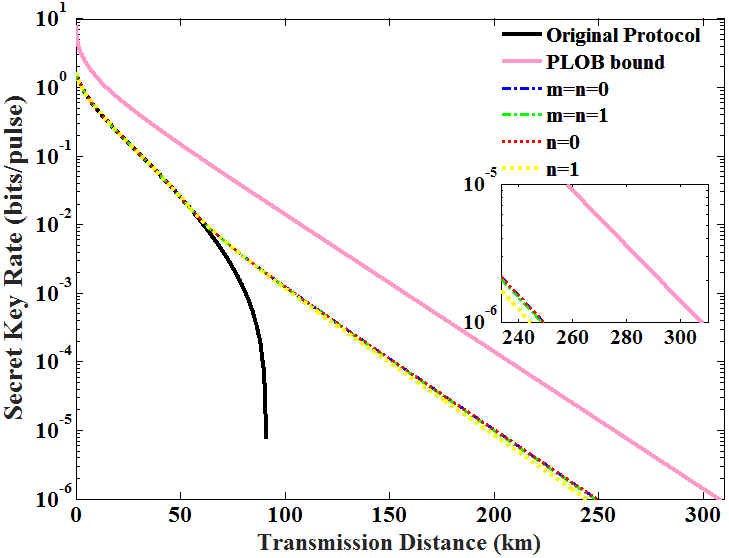}
\label{Fig8a}
}
\subfigure[]{
\includegraphics[width=\columnwidth]{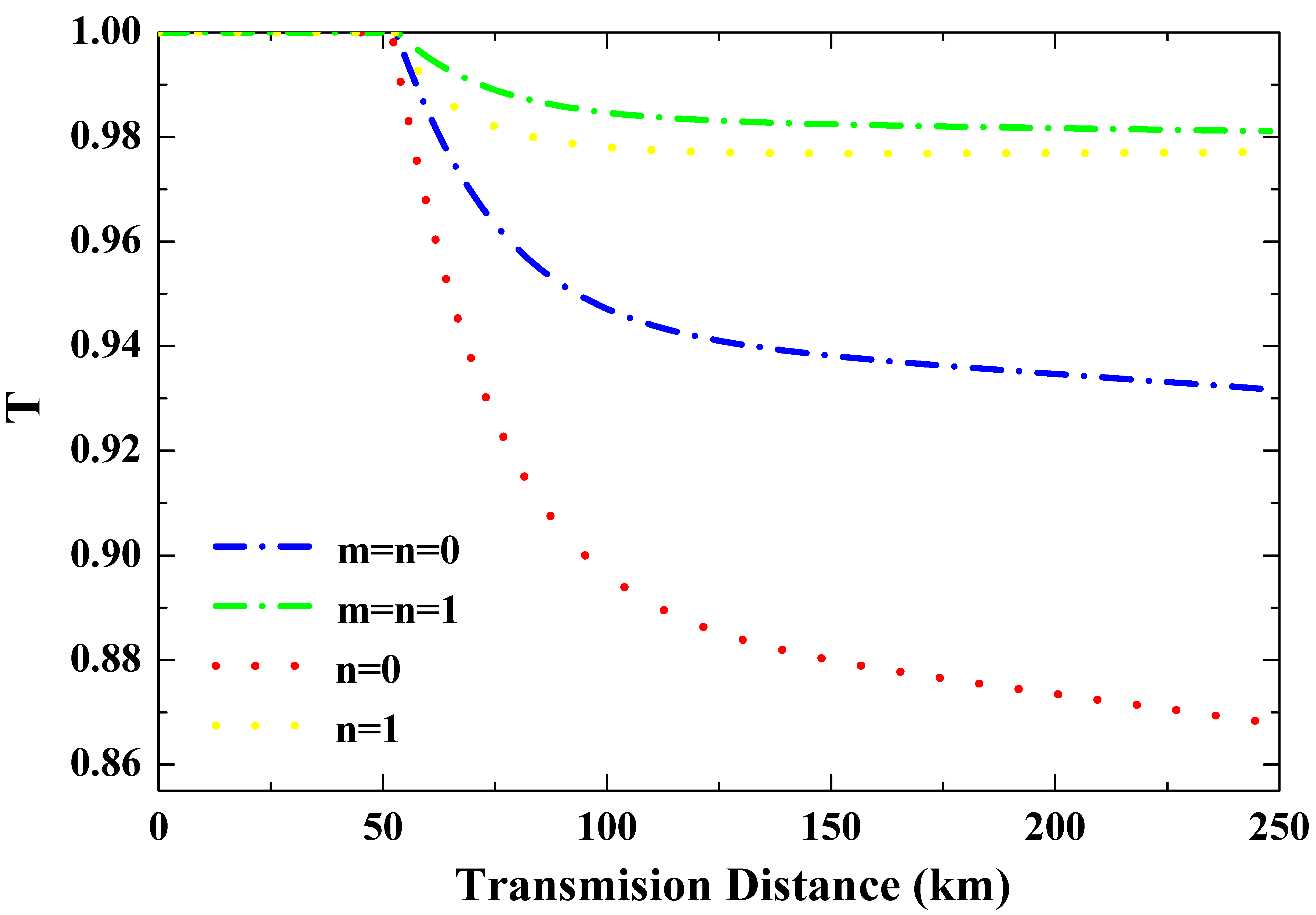}
\label{Fig8b}
} \newline
\newline
\caption{(Color online) (a) The secret key rates of the QC-CVQKD system
[dash-dotted lines and dotted lines] as a function of the transmission
distance for bilateral symmetrical quantum catalysis (BSQC) cases ($%
T_{1}=T_{2}=T$ and $m=n\in\{0,1\}$) [dash-dotted line] and single side
quantum catalysis (SSQC) cases ($T_{1}=1$, $T_{2}=T$ and $n\in\{0,1\}$)
[dotted line] for the optimal choice of $T$. (b) The secret key rates of the
QC-CVQKD system with the optimal $T$. The variance of EPR state is $V=20$,
the excess noise is $\protect\varepsilon=0.01$, and the reconciliation
efficiency is $\protect\beta=0.95$. }
\end{figure}

\subsection{Security analysis}

To evaluate the performance of QC-CVQKD system, according to the detailed
calculations of asymptotic secret key rate [see in Appendix B], we
demonstrate the numerical simulations of the secret key rate and tolerable
excess noise.

Fig.7 shows that for a given transmittance $T=0.95$, the asymptotic secret
key rate $\widetilde{K}_{R}$ as a function of transmission distance can be
plotted with different photon-catalyzed numbers $m=n\in\{0,1,2\}$ and $%
n\in\{0,1,2\}$. The black solid line denotes the secret key rate of the
original protocol, which is exceeded by the QC-CVQKD system with zero-photon
and single-photon quantum catalysis within the long-distance range. To be
specific, the proposed system of using zero-photon BSQC [blue dash-dotted
line] has the longer transmission distance when compared with the
zero-photon SSQC case [red dotted line]. While for the single-photon
QC-CVQKD system, the BSQC [green dash-dotted line] in term of the maximum
transmission distance is better than the SSQC case [yellow dotted line]. The
reason may be that for the single-photon BSQC, extra adding the model of
quantum catalysis $\hat{O}_{m}$ before Alice taking heterodyne detection can
be regarded as the generation of trusted noise controlled by Alice, thereby
resulting in the diminution of the Holevo bound $S^{G}\left( B\text{:}%
E\right) $. However, for the two-photon QC-CVQKD system, the BSQC [cyan
dash-dotted line] and SSQC [orange dotted line] are worse than the original
one, resulted from the fact that the more photons are catalyzed, the higher
the non-Gaussianity is, thereby making the more noise to the covariance
matrix \cite{42,43}. In addition, Within the shortening distance, the secret
key of the QC-CVQKD system is worse than that of the original system because
of the limitation of the success probability of quantum catalysis. As a
result, for a given large transmittance $T=0.95$, the QC-CVQKD system of
using zero-photon and single- photon quantum catalysis can lengthen the
maximal transmission distance apart from the two-photon QC-CVQKD system.

Since it is so, for the optimal choice of $T$, we obtain the maximal secret
key rate of the proposed system with zero-photon and single-photon quantum
catalysis. In Fig.8, we show the maximal secret key rate as a function of
transmission distance for $m=n\in\{0,1\}$ and $n\in\{0,1\}$, when compared
with the original protocol [black solid line]. In Fig.8(b), it is a case of
the optimal $T$ that achieves the maximal secret key rate. We find that, for
the long-distance range, the zero-photon and single-photon QC-CVQKD systems
at the optimal transmittance range ($0.86\leqslant T\leqslant1$) perform
better than the original system, in terms of both secret key rate and
transmission distance. It indicates that the quantum catalysis can be used
for improving the performance of CVQKD. For the single-photon QC-CVQKD
system [green dash-dotted line and yellow dotted line] at the long
transmission distance, the range of the optimal $T$ is approximate $%
0.978\leqslant T\leqslant1$ in which there does exist a high success
probability for single-photon quantum catalysis [see Fig.6a]. However, for
the short-distance range, even if for the optimal choice of $T$, the secret
key rate of the QC-CVQKD system is similar to that of the original system,
because for $T_{1}=T_{2}=1$ of Alice$^{\text{'}}$s BS$_{\text{I}}$ and BS$_{%
\text{II}}$, there is no quantum catalysis effect resulting from the CVQKD
system.

\begin{figure}[ptb]
\label{Fig.9} \centering \includegraphics[width=\columnwidth]{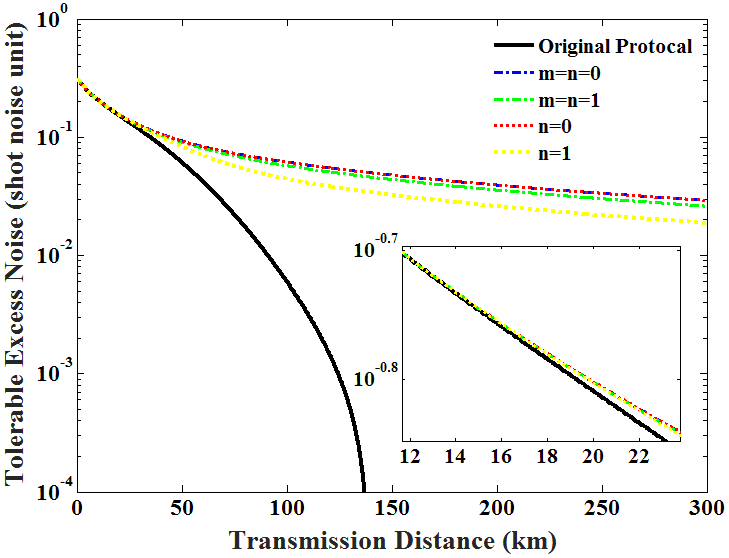}
\caption{(Color online) The maximal tolerable excess noise of the QC-CVQKD
system [dash-dotted lines and dotted lines] as a function of the
transmission distance for bilateral symmetrical quantum catalysis (BSQC)
cases ($T_{1}=T_{2}=T$ and $m=n\in\{0,1\}$) [dash-dotted line] and single
side quantum catalysis (SSQC) cases ($T_{1}=1$, $T_{2}=T$ and $n\in\{0,1\}$)
[dotted line] for the optimal choice of $T$. The variance of EPR state is $%
V=20$ and the reconciliation efficiency is $\protect\beta=0.95$. }
\end{figure}

\begin{figure}[ptb]
\label{Fig.10} \centering
\subfigure[]{
\includegraphics[width=\columnwidth]{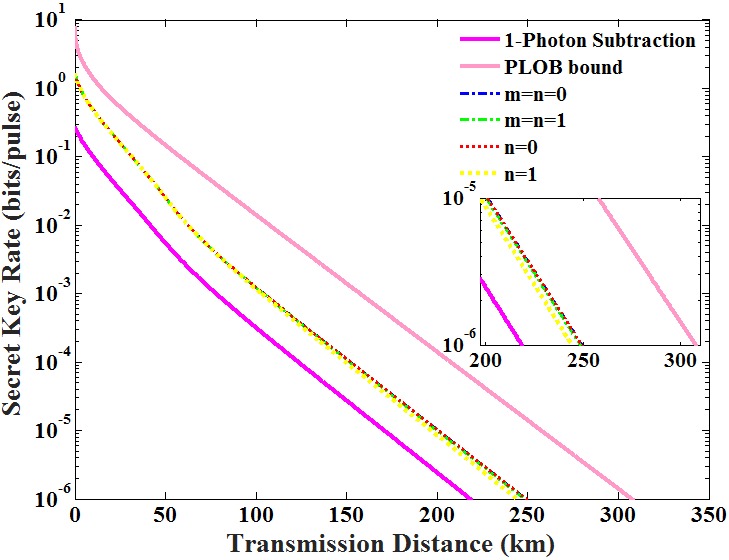}
\label{Fig10a}
}
\subfigure[]{
\includegraphics[width=\columnwidth]{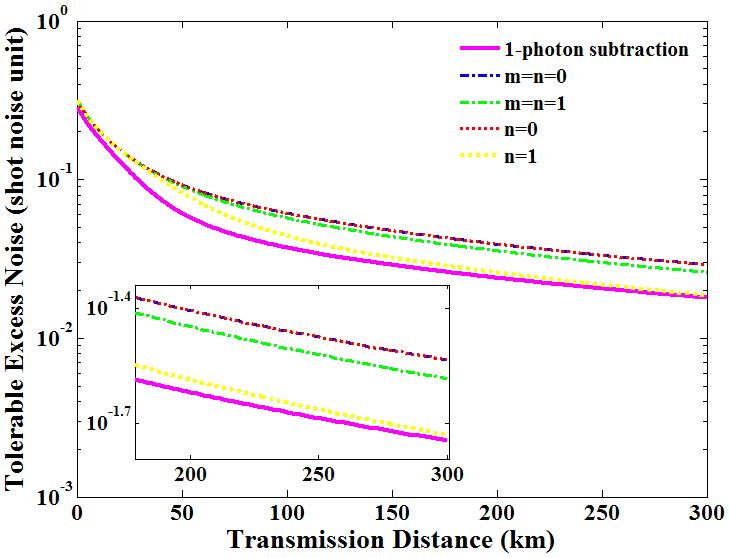}
\label{Fig10b}
} \newline
\newline
\caption{(Color online) (a) The maximal secret key rate for $\protect\epsilon%
=0.01$ and (b) the maximal tolerable excess noise of the QC-CVQKD system
[dash-dotted lines and dotted lines] and the CVQKD with the single-photon
subtraction [magenta solid line] as a function of the transmission distance
for bilateral symmetrical quantum catalysis (BSQC) cases ($T_{1}=T_{2}=T$
and $m=n=0,1$) [dash-dotted line] and single side quantum catalysis (SSQC)
cases ($T_{1}=1$, $T_{2}=T$ and $n=0,1)$ [dotted line] for the optimal
choice of $T$. The variance of EPR state is $V=20$ and the reconciliation
efficiency is $\protect\beta=0.95$. }
\end{figure}

Additionally, the other factor that has an effect on the QC-CVQKD system is
the tolerable excess noise. In Fig.9, we illustrate the tolerable excess
noise as a function of transmission distance for the optimal choice of $T$.
Analogous to Fig.8, at the long-distance range, the QC-CVQKD system with
zero-photon and single-photon quantum catalysis exceed the original system
with respect to the maximal tolerable excess noise for remote users. More
specifically, the zero-photon QC-CVQKD system [blue dash-dotted line and red
dotted line] presents the best performance since the maximal tolerable
excess noise approaches to about $0.0292$ at the transmission distance of $%
300$ km. Besides, at the transmission distance of $300$ km, for the
single-photon BSQC (i.e.,$m=n=1$) [green dash-dotted line] and SSQC (i.e.,$%
n=1$) [yellow dotted line], the maximal tolerable excess noises can approach
to about $0.0261$ and $0.0185$, respectively. There results indicate that
when the quantum channel is less noise ($\varepsilon \sim 0.0185$), the
zero-photon and single-photon quantum catalysis can be applied to lengthen
the maximal transmission distance up to hundreds of kilometers. In addition,
we find that from Fig.8(a) and Fig.9, at the long distance range, for the
single-photon QC-CVQKD schemes, the BSQC case [green dash-dotted line]
performs better than the SSQC [yellow dotted line]. It indicates that the
single-photon QC-CVQKD system, extra adding the model of quantum catalysis $%
\hat{O}_{m}$ in mode A may be useful for improving the performance of CVQKD
protocol, when compared with the SSQC $\hat{O}_{n}$ in mode B.

Interestingly,\ in Ref.\cite{11}, it was pointed out that for the
photon-subtraction-involved CVQKD system, the single-photon subtraction
operation can usually improve the performance of the related system.
Therefore, in order to make comparisons of the QC-CVQKD and the
single-photon-subtraction(SS)-CVQKD, here we give the schematic diagram of
the SS-CVQKD system in Fig.11. We consider the asymptotic secret key rate $%
K_{asy}$ of reverse reconciliation under collective attack with the
assistant of the IWOP technique [The more detailed calculations can be seen
in Appendix C]. To display the effect of quantum catalysis on the
performance of CVQKD, we plot the secret key rate and the tolerable excess
noise of the CVQKD system involving quantum catalysis and single-photon
subtraction as a function of transmission distance for photon-catalyzed
numbers $m=n\in \{0,1\}$ and $n\in \{0,1\}$, as shown in Fig.10(a) and
Fig.10(b), respectively. It is found that the performance of the SS-CVQKD
system [magenta solid line] in terms of the maximal secret key rate and the
maximal tolerable excess noise is outperformed by the QC-CVQKD system at the
long transmission distance range. The reason may be that the success
probability for single-photon subtraction is lower than that for quantum
catalysis at the optimal transmittance $T$ range [see Figs.8(b)]. It implies
that the former loses more information than the latter in the process of
distilling a common secret key. Without loss of generality, we assume that
the minimal secret key rate is confined to above $10^{-6}$ bits per pulse.
For the single-photon QC-CVQKD system [green dash-dotted line and yellow
dotted line], for the optimal choice of the transmittance $T$ of Alice$^{%
\text{'}}$s beamsplitters, the maximal transmission distances is more than $%
240$ km. Whereas for the SS-CVQKD system, the maximal transmission distance
is approximate $218$ km, because its success probability is limited to below
$0.25$ [magenta surface in Fig.6(a)]. These comparison results show that the
performance of the QC-CVQKD system using zero-photon and single-photon
quantum catalysis performs better than the SS-CVQKD system when optimized
over the transmittance $T$.

Attractively, from Figs.7, 8(a) and 10(a), we also consider the PLOB bound
that stands for the fundamental rate-loss scalling (secret key capacity)
\cite{45}. By comparison, it is found that for a given transmittance $T=0.95$%
, the performance of QC-CVQKD system using\ the zero-photon BSQC (i.e.,$m=n=0
$) is closer to the PLOB bound than that of original protocol when the
transmission distance reaches larger than $57.6851$ km. While for the
optimal choice of the transmittance $T$, we can easily see that, our
proposed QC-CVQKD system involving the zero-photon and single-photon quantum
catalysis is closer to the PLOB bound when comparing with the SS-CVQKD.
However, both of them are unable to exceed the PLOB bound at any
transmission distance. Therefore, in order to beat the PLOB bound that is
ultimate limit of repeaterless point-to-point communication, we can design
the one way continuous-variable measurement-device-independent QC-QKD system
acting as an active repeater.

\section{Conclusion}

We have suggested the effect of quantum catalysis on the performance of the
CVQKD system by using the IWOP technique. From the equivalent operator of
quantum catalysis, the quantum catalysis that is a non-Gaussian operation in
essence can be used for improving the CVQKD system. Different from the
traditional TMSV, the entanglement of the resulted state using quantum
catalysis can be improved significantly after optimizing the transmittance $%
T $ of Alice beamsplitters, and the success probability for quantum
catalysis in high transmittance $T$ performs better than the single-photon
subtraction case, especially for the zero-photon quantum catalysis. Taking
into account the Gaussian optimality, we derive the lower bound of the
asymptotic secret key rate of the QC-CVQKD for reverse reconciliation
against the collective attack. Numerical simulations show that when
comparing with the SS-CVQKD system, the QC-CVQKD system has an advantage of
lengthening the maximal transmission distance with the raised secret key
rates. For all the QC-CVQKD systems, the zero-photon quantum catalysis has
the best performance. While for the QC-CVQKD system using single-photon
quantum catalysis, the BSQC performs better than the SSQC due to the fact
that extra adding the model of quantum catalysis $\hat{O}_{m}$ is useful for
improving the performance of the CVQKD system. We make a comparison of the
CVQKD systems involving quantum catalysis and single-photon subtraction. It
is found that the QC-CVQKD system using the zero-photon and single-photon
quantum catalysis is superior to the single-photon subtraction case in terms
of the maximal transmission distance.

\begin{acknowledgments}
We would like to thank Professor S. Pirandola for his helpful suggestion. This work was supported by the National Natural Science Foundation of China
(Grant Nos. 61572529, 61871407).
\end{acknowledgments}

\begin{appendix}
\section{Derivation of the success probability }$P_{d}$

In order to derive the analytical expression of the success probability%
\textbf{\ }$P_{d}$ shown in Eq.(\ref{10}), we rewrite the state in Eq.(\ref%
{9}) as the density operator $\rho =\left\vert \psi \right\rangle
\left\langle \psi \right\vert $, i.e.,
\begin{align}
\rho _{A_{1}B_{1}}& =\frac{1}{P_{d}}\hat{O}_{m}\hat{O}_{n}\left\vert
TMSV\right\rangle \left\langle TMSV\right\vert \hat{O}_{n}^{\dag }\hat{O}%
_{m}^{\dag }  \notag \\
& =\frac{W_{0}^{2}}{P_{d}}\Re ^{m,n}\Pi \exp [a^{\dag }b^{\dag }W]00\rangle
\left\langle 00\right\vert \exp [abW_{1}]  \tag{A1}
\end{align}%
where we have used the equivalent operators of photon catalysis operations
in Eq.(\ref{4}), e$^{\zeta a^{\dag }a}a^{\dag }$e$^{-\zeta a^{\dag
}a}=a^{\dag }e^{\zeta },$ and set
\begin{align}
& W=\frac{\lambda \left( T_{2}-\gamma \right) \left( T_{1}-\tau \right) }{%
\sqrt{T_{1}T_{2}}\left( 1-\gamma \right) \left( 1-\tau \right) },  \notag \\
& W_{0}=\frac{\sqrt{T_{1}^{m}T_{2}^{n}\left( 1-\lambda ^{2}\right) }}{n!m!},
\notag \\
& W_{1}=\frac{\lambda \left( T_{2}-\gamma _{1}\right) \left( T_{1}-\tau
_{1}\right) }{\sqrt{T_{1}T_{2}}\left( 1-\gamma _{1}\right) \left( 1-\tau
_{1}\right) },  \notag \\
& \Pi =\frac{1}{1-\tau }\frac{1}{1-\gamma }\frac{1}{1-\tau _{1}}\frac{1}{%
1-\gamma _{1}},  \notag \\
& \Re ^{m,n}=\frac{\partial ^{m}}{\partial \tau ^{m}}\frac{\partial ^{n}}{%
\partial \gamma ^{n}}\frac{\partial ^{m}}{\partial \tau _{1}^{m}}\frac{%
\partial ^{n}}{\partial \gamma _{1}^{n}}\left\{ ...\right\} |_{\tau =\gamma
=\tau _{1}=\gamma _{1}=0}.  \tag{A2}
\end{align}%
Then, according to the completeness relation of coherent state
representation $\int d^{2}z\left\vert z\right\rangle \left\langle
z\right\vert /\pi =1$,\ the integrational formula%
\begin{align}
& \int \frac{d^{2}z}{\pi }\exp \left( \zeta \left\vert z\right\vert ^{2}+\xi
z+\eta z^{\ast }+fz^{2}+gz^{\ast 2}\right)  \notag \\
& =\frac{1}{\sqrt{\zeta ^{2}-4fg}}\exp \left[ \frac{-\zeta \xi \eta +\xi
^{2}g+\eta ^{2}f}{\zeta ^{2}-4fg}\right] ,  \tag{A3}
\end{align}%
and the completeness of the resulted state Tr($\rho _{A_{1}B_{1}}^{N}$)=1,
we can obtain the success probability\textbf{\ }$P_{d}$ given by%
\begin{align}
P_{d}& =W_{0}^{2}\Re ^{m,n}\Pi \left\langle 00\right\vert \exp [abW_{1}]\exp
[a^{\dag }b^{\dag }W]00\rangle  \notag \\
& =W_{0}^{2}\Re ^{m,n}\Pi \int \frac{d^{2}z}{\pi ^{2}}\int \frac{d^{2}\beta
}{\pi ^{2}}\exp \left[ -\left\vert z\right\vert ^{2}\right.  \notag \\
& \left. -\left\vert \beta \right\vert ^{2}+z\beta W_{1}+z^{\ast }\beta
^{\ast }W\right]  \notag \\
& =W_{0}^{2}\Re ^{m,n}\left\{ \frac{\Pi }{1-W_{1}W}\right\} .  \tag{A4}
\end{align}

\section{Calculation of asymptotic secret key rate }

Here, we present the calculation of asymptotic secret key rates of the
QC-CVQKD system where Alice performs heterodyne detection and Bob performs
homodyne detection. As mentioned above, state $\left \vert \psi
\right
\rangle _{A_{1}B_{1}}$ belongs to a new kind of non-Gaussian state,
thus we cannot directly use the results of the conventional Gaussian CVQKD
to calculate its secret key rate. Fortunately, thanks to the extremity of
Gaussian quantum states that the rendering secret key rate of the
non-Gaussian state $\left \vert \psi \right \rangle _{A_{1}B_{1}}$ is no
less than that of a Gaussian state $\left \vert \psi \right \rangle
_{A_{1}B_{1}}^{G}$ with the same covariance matrix $\Gamma_{A_{1}B_{1}}=$ $%
\Gamma_{A_{1}B_{1}}^{G}$, we obtain $K$($\left \vert \psi \right \rangle
_{A_{1}B_{1}}$)$\geqslant K$($\left \vert \psi \right \rangle
_{A_{1}B_{1}}^{G}$) \cite{16,18}. For reverse reconciliation, therefore, the
lower bound of the asymptotic secret key rate under optimal collective
attack can be given by%
\begin{equation}
\widetilde{K}_{R}=P_{d}\left \{ \beta I^{G}\left( A:B\right) -S^{G}\left(
B:E\right) \right \} ,  \tag{B1}
\end{equation}
where $\beta$ denotes the reconciliation efficiency, $I^{G}\left( A\text{:}%
B\right) $ denotes Alice and Bob$^{\text{'}}$s mutual information, and $%
S^{G}\left( B\text{:}E\right) $ denotes the Holevo bound, which is defined
as the maximum information on Bob final key available to Eve.

In order to derive the analytical expression of the asymptotic secret key
rate $K$($\left \vert \psi \right \rangle _{A_{1}B_{1}}^{G}$), we consider
the covariance matrix $\Gamma_{A_{1}B_{1}}$ of the resulted state $%
\left
\vert \psi \right \rangle _{A_{1}B_{1}}$ given by
\begin{equation}
\Gamma_{A_{1}B_{1}}=\left(
\begin{array}{cc}
X_{A}II & Z_{AB}\sigma_{z} \\
Z_{AB}\sigma_{z} & Y_{B}II%
\end{array}
\right) ,  \tag{B2}
\end{equation}
where $II=diag(1,1)$, $\sigma_{z}=diag(1,-1)$, and $X_{A},Y_{B}$ and $Z_{AB}
$ can be derived by using the IWOP technique as follows: It is first
required to derive the average values such as $\left \langle
a^{\dagger}a\right \rangle ,\left \langle b^{\dagger}b\right \rangle $ and $%
\left \langle ab\right \rangle $. According to Eq.(A1) and\ Eq.(A3), thus,
it is straightforward to get

\begin{align}
\left \langle a^{\dagger}a\right \rangle & =\text{Tr[}\rho_{A_{1}B_{1}}^{N}%
\left( aa^{\dagger}-1\right) \text{]}  \notag \\
& =\frac{W_{0}^{2}}{P_{d}}\Re^{m,n}\Pi \int \frac{d^{2}\alpha}{\pi^{2}}\int
\frac{d^{2}\beta}{\pi^{2}}\alpha \alpha^{\ast}  \notag \\
& \times \exp[-\left \vert \alpha \right \vert ^{2}-\left \vert \beta \right
\vert ^{2}+\alpha \beta W_{1}+\alpha^{\ast}\beta^{\ast}W]-1  \notag \\
& =\frac{W_{0}^{2}}{P_{d}}\Re^{m,n}\left \{ \frac{\Pi}{\left(
1-W_{1}W\right) ^{2}}\right \} -1,  \tag{B3}
\end{align}%
\begin{align}
\left \langle b^{\dagger}b\right \rangle & =\text{Tr[}\rho_{A_{1}B_{1}}^{N}%
\left( bb^{\dagger}-1\right) \text{]}  \notag \\
& =\frac{W_{0}^{2}}{P_{d}}\Re^{m,n}\Pi \int \frac{d^{2}\alpha}{\pi^{2}}\int
\frac{d^{2}\beta}{\pi^{2}}\beta \beta^{\ast}  \notag \\
& \times \exp[-\left \vert \alpha \right \vert ^{2}-\left \vert \beta \right
\vert ^{2}+\alpha \beta W_{1}+\alpha^{\ast}\beta^{\ast}W]-1  \notag \\
& =\frac{W_{0}^{2}}{P_{d}}\Re^{m,n}\left \{ \frac{\Pi}{\left(
1-W_{1}W\right) ^{2}}\right \} -1,  \notag \\
& =\left \langle a^{\dagger}a\right \rangle ,  \tag{B4}
\end{align}%
\begin{align}
\left \langle ab\right \rangle & =\text{Tr[}\rho_{A_{1}B_{1}}^{N}ab\text{]}
\notag \\
& =\frac{W_{0}^{2}}{P_{d}}\Re^{m,n}\Pi \int \frac{d^{2}\alpha}{\pi^{2}}\int
\frac{d^{2}\beta}{\pi^{2}}\alpha \beta  \notag \\
& \times \exp[-\left \vert \alpha \right \vert ^{2}-\left \vert \beta \right
\vert ^{2}+\alpha \beta W_{1}+\alpha^{\ast}\beta^{\ast}W]  \notag \\
& =\frac{W_{0}^{2}}{P_{d}}\Re^{m,n}\left \{ \frac{\Pi W}{\left(
1-W_{1}W\right) ^{2}}\right \} .  \tag{B5}
\end{align}
Note that $\left \langle ab\right \rangle =\left \langle
a^{\dagger}b^{\dagger }\right \rangle ^{\dagger}.$By combining
Eqs.(B3)-(B5), therefore, we can directly obtain the elements of convariance
matrix $\Gamma_{A_{1}B_{1}}^{N}$ as the following form

\begin{align}
X_{A} & =\text{Tr}\left[ 1+2a^{\dagger}a\right]  \notag \\
& =\frac{2W_{0}^{2}}{P_{d}}\Re^{m,n}\left \{ \frac{\Pi}{\left(
1-W_{1}W\right) ^{2}}\right \} -1,  \notag \\
Y_{B} & =\text{Tr}\left[ 1+2b^{\dagger}b\right] =X_{A},  \notag \\
Z_{AB} & =\text{Tr}\left[ ab+a^{\dagger}b^{\dagger}\right]  \notag \\
& =\frac{2W_{0}^{2}}{P_{d}}\Re^{m,n}\left \{ \frac{\Pi W}{\left(
1-W_{1}W\right) ^{2}}\right \} .  \tag{B6}
\end{align}

After passing the untrusted quantum channel which is characterized by the
transmission efficiency $T_{c}$ and the excess noise $\varepsilon$, the
covariance matrix $\Gamma_{A_{1}B_{2}}^{G}$ reads

\begin{equation}
\Gamma_{A_{1}B_{2}}^{G}=\left(
\begin{array}{cc}
X_{A}II & \sqrt{T_{c}}Z_{AB}\sigma_{z} \\
\sqrt{T_{c}}Z_{AB}\sigma_{z} & T_{c}\left( X_{A}+\xi \right) II%
\end{array}
\right) ,  \tag{B7}
\end{equation}
where $\xi=\left( 1-T_{c}\right) /T_{c}+\varepsilon$ denotes the
channel-added noise referred to the input of Gaussian channel. Tthe mutual
information between Alice and Bob now can be expressed as%
\begin{align}
I^{G}\left( A\text{:}B\right) & =\frac{1}{2}\log_{2}\frac{V_{A_{1}}}{%
V_{A_{1}|B_{2}}}  \notag \\
& =\log_{2}\left \{ \sqrt{\frac{\left( X_{A}+1\right) \left( X_{A}+\xi
\right) }{\left( X_{A}+1\right) \left( X_{A}+\xi \right) -Z_{AB}^{2}}}\right
\} .  \tag{B8}
\end{align}
Furthermore, Eve accessible quantum information on Bob measurement can be
calculated by assuming Eve can purify the whole system $S^{G}\left( B\text{:}%
E\right) =S\left( E\right) -S\left( E|B\right) =S\left( AB\right) -S\left(
A|B\right) $. For the Gaussian modulation, the first term $S\left( AB\right)
$ is a function of the symplectic eigenvalues $\lambda_{1,2}$ of $%
\Gamma_{A_{1}B_{2}}^{G}$, which is given by
\begin{subequations}
\begin{equation}
S\left( AB\right) =G\left[ \left( \lambda_{1}-1\right) /2\right] +G\left[
\left( \lambda_{2}-1\right) /2\right] ,  \tag{B9}
\end{equation}
where the Von Neumann entropy $G\left[ x\right] $ is
\end{subequations}
\begin{equation}
G\left[ x\right] =\left( x+1\right) \log_{2}\left( x+1\right) -x\log _{2}x
\tag{B10}
\end{equation}
and
\begin{equation}
\lambda_{1,2}^{2}=\frac{1}{2}\left[ \Lambda \pm \sqrt{\Lambda^{2}-4D^{2}}%
\right] ,  \tag{B11}
\end{equation}
with the notation%
\begin{align}
\Lambda & =X_{A}^{2}+T_{c}^{2}\left( X_{A}+\xi \right) ^{2}-2T_{c}Z_{AB}^{2},
\notag \\
D & =X_{A}T_{c}\left( X_{A}+\xi \right) -T_{c}Z_{AB}^{2}.  \tag{B12}
\end{align}

Moreover, the second term $S\left( A|B\right) =G\left[ \left( \lambda
_{3}-1\right) /2\right] $ is a function of the symplectic eigenvalue $%
\lambda_{3}$ of the covariance matrix $\Gamma_{A}^{b}$ of Alice mode after
Bob performing homodyne detection, where the square of symplectic eigenvalue
$\lambda_{3}$ is%
\begin{equation}
\lambda_{3}^{2}=X_{A}\left[ X_{A}-\frac{Z_{AB}^{2}}{X_{A}+\xi}\right] .
\tag{B13}
\end{equation}

As a result, the asymptotic secret key rate can be written as
\begin{equation}
\widetilde{K}_{R}=P_{d}\left \{ \beta I^{G}\left( A:B\right) -S\left(
AB\right) +S\left( A|B\right) \right \} .  \tag{B14}
\end{equation}
\

\begin{figure}[ptb]
\label{Fig11} \centering \includegraphics[width=\columnwidth]{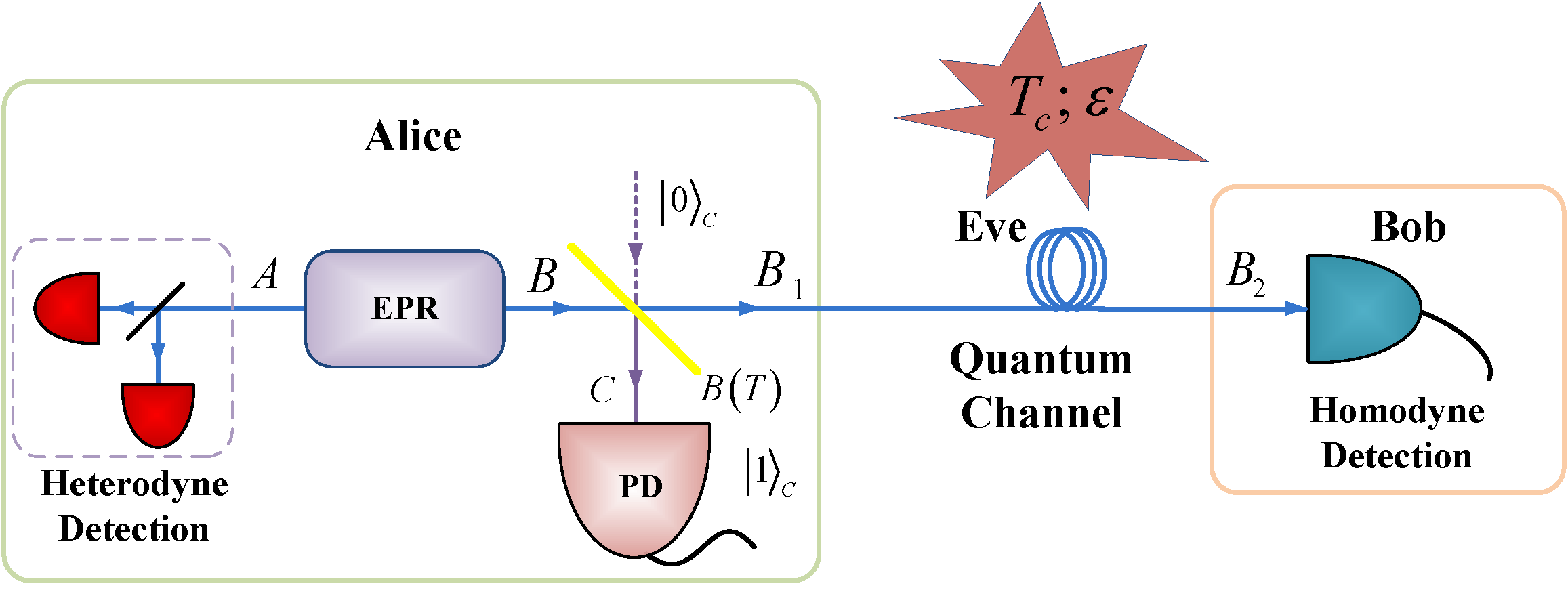} \ \
\ \ \ \
\caption{(Color online) The schematic diagram of the Gaussian modulation
CVQKD scheme with single-photon subtraction.}
\end{figure}

\section{The secret key rate of the \textbf{single-photon
subtraction }EB-CVQKD protocol under collective attack }

In order to make a comparison of the proposed long-distance CVQKD scheme via
quantum catalysis, here we review the CVQKD protocol of applying single
photon subtraction, and then assume that these two schemes have the same
quantum channel controlled by Eve. As can be seen from Fig.11, Alice
generates a two-mode squeezed vacuum state $\left \vert TMSV\right \rangle
_{AB}$ (EPR), and performs heterodyne detection of the one half of $%
\left
\vert TMSV\right \rangle _{AB}$. The other half of $\left \vert
TMSV\right
\rangle _{AB}$ after operating single-photon subtraction is sent
to Bob through the same quantum channel marked by transmission efficiency $%
T_{c}$ and excess noise $\varepsilon$. Afterwards, Bob performs homodyne
detection of the received state and then informs Alice about which
observable he measured, so that two correlated variables, which are shared
by both Alice and Bob, can be used to exact a common secret key.

In deed, starting from the concept of quantum operators, the single-phton
subtraction operation can be seen as an equivalent operator $\Theta $ which
is given by
\begin{align}
\Theta & \text{=}_{C}\left\langle 1\right\vert B\left( T\right) \left\vert
0\right\rangle _{C}  \notag \\
& =\frac{1-T}{T}b\exp \left[ b^{\dagger }b\ln \sqrt{T}\right]  \tag{C1}
\end{align}%
Thus, the photon-subtraction state $\left\vert \Psi \right\rangle _{AB_{1}}$
after operating single-photon subtraction is expressed as%
\begin{align}
\left\vert \Psi \right\rangle _{AB_{1}}& =\frac{1}{\sqrt{P_{1}}}\Theta
\left\vert TMSV\right\rangle _{AB}  \notag \\
& =\frac{\widetilde{A}\widetilde{B}}{\sqrt{P_{1}}}\exp \left[ \widetilde{B}%
a^{\dagger }b^{\dagger }\right] a^{\dagger }\left\vert 00\right\rangle _{AB}
\tag{C2}
\end{align}%
where
\begin{align}
\widetilde{A}& =\sqrt{\frac{\left( 1-\lambda ^{2}\right) \left( 1-T\right) }{%
T}},  \notag \\
\widetilde{B}& =\lambda \sqrt{T},  \tag{C3}
\end{align}%
and
\begin{equation}
P_{1}=\frac{\widetilde{A}^{2}\widetilde{B}^{2}}{\left( 1-\widetilde{B}%
^{2}\right) ^{2}}  \tag{C4}
\end{equation}%
is the success probability of implementing the single-photon subtraction
operation. After the photon-subtraction state $\left\vert \Psi \right\rangle
_{AB_{1}}$ goes through the quantum channel, similarly to Eq.(B7), we also
can obtain the covariance matrix $\Gamma ^{1}$ as the following form
\begin{equation}
\Gamma ^{1}=\left(
\begin{array}{cc}
XII & \sqrt{T_{c}}Z\sigma _{z} \\
\sqrt{T_{c}}Z\sigma _{z} & T_{c}\left( Y+\xi \right) II%
\end{array}%
\right)  \tag{C5}
\end{equation}%
where $\xi =\left( 1-T_{c}\right) /T_{c}+\varepsilon $, and
\begin{align}
X& =\frac{4\widetilde{A}^{2}\widetilde{B}^{2}}{P_{1}\left( 1-\widetilde{B}%
^{2}\right) ^{3}}-1,  \notag \\
Y& =\frac{2\widetilde{A}^{2}\widetilde{B}^{2}\left( 1+\widetilde{B}%
^{2}\right) }{P_{1}\left( 1-\widetilde{B}^{2}\right) ^{3}}-1,  \notag \\
Z& =\frac{4\widetilde{A}^{2}\widetilde{B}^{3}}{P_{1}\left( 1-\widetilde{B}%
^{2}\right) ^{3}}.  \tag{C6}
\end{align}

Now let us consider the calculation of asymptotic secret key rate of the
single-photon subtraction EB-CVQKD protocol in the context of Gaussian
optimality theorem. Thus, the lower bound of asymptotic secret key rate $%
K_{asy}$ of reverse reconciliation under collective attack is
\begin{equation}
K_{asy}=P_{1}\left \{ \beta I^{Hom}\left( A\text{:}B\right) -S^{Hom}\left( B%
\text{:}E\right) \right \}  \tag{C7}
\end{equation}
where $P_{1}$ has been derived in Eq.(C4), $\beta$ is the efficiency for
reverse reconciliation, and the superscript Hom represents Bob taking
homodyne detection. Additionally, the mutual information between Alice and
Bob is given by
\begin{equation}
I^{Hom}\left( A\text{:}B\right) =\log_{2}\left \{ \sqrt{\frac{\left(
X+1\right) \left( Y+\xi \right) }{\left( X+1\right) \left( Y+\xi \right)
-Z^{2}}}\right \}  \tag{C8}
\end{equation}

Under the assumption that Eve is able to purity the whole system, $%
S^{G}\left( B\text{:}E\right) =S\left( E\right) -S\left( E|B\right) =S\left(
AB\right) -S\left( A|B\right) $, we can directly obtain the symplectic
eigenvalues $\widetilde{\lambda}_{1,2,}$ of covariance matrix $\Gamma^{1}$
as the following form%
\begin{equation}
\widetilde{\lambda}_{1,2}^{2}=\frac{1}{2}\left[ \widetilde{C}\pm \sqrt{%
\widetilde{C}^{2}-4\widetilde{D}^{2}}\right] ,  \tag{C9}
\end{equation}
with%
\begin{align}
\widetilde{C} & =X^{2}+T_{c}^{2}\left( Y+\xi \right) ^{2}-2T_{c}Z^{2}  \notag
\\
\widetilde{D} & =XT_{c}\left( Y+\xi \right) -T_{c}Z^{2}  \tag{C10}
\end{align}
and
\begin{equation}
\widetilde{\lambda}_{3}^{2}=X\left[ X-\frac{Z^{2}}{Y+\xi}\right] .  \tag{C11}
\end{equation}
Furthermore, $S\left( AB\right) =G\left[ \left( \widetilde{\lambda}%
_{1}-1\right) /2\right] +G\left[ \left( \widetilde{\lambda}_{2}-1\right) /2%
\right] $ and $S\left( A|B\right) =G\left[ \left( \widetilde{\lambda }%
_{3}-1\right) /2\right] $ where the Von Neumann entropy $G\left[ x\right] $
is defined in Eq.(B10).
\end{appendix}

\end{document}